\documentclass[aps,amsmath,amssymb,superscriptaddress,prb,reprint,twocolumn]{revtex4-2}
\usepackage{graphicx}% Include figure files
\usepackage{dcolumn}% Align table columns on decimal point
\usepackage{bm}% bold math
\usepackage{here}
\usepackage{mathtools}
\usepackage{braket}

\begin{document}

\preprint{APS/123-QED}

\title{Strain Effects on Electronic Properties of Cobalt-Based Coordination Nanosheets}
\author{Kento Nishigomi}
\affiliation{Department of Nanotechnology for Sustainable
 Energy, School of Science and Technology, Kwansei Gakuin University, Gakuen-Uegahara 1, Sanda 669-1330, Japan}
\author{Yu Yi}
\affiliation{Department of Nanotechnology for Sustainable
 Energy, School of Science and Technology, Kwansei Gakuin University, Gakuen-Uegahara 1, Sanda 669-1330, Japan}
\author{Souren Adhikary}
\affiliation{Department of Nanotechnology for Sustainable
 Energy, School of Science and Technology, Kwansei Gakuin University, Gakuen-Uegahara 1, Sanda 669-1330, Japan}
\author{Kazuhito Tsukagoshi}
\affiliation{Research Center for Materials Nanoarchitectonics (MANA), National Institute for Materials Science (NIMS), Namiki 1-1, Tsukuba 305-0044, Japan}
\author{Katsunori Wakabayashi}
\affiliation{Department of Nanotechnology for Sustainable
 Energy, School of Science and Technology, Kwansei Gakuin University, Gakuen-Uegahara 1, Sanda 669-1330, Japan}
\affiliation{National Institute for Materials Science (NIMS), Namiki 1-1, Tsukuba 305-0044, Japan}
\affiliation{Center for Spintronics Research Network (CSRN), Osaka
University, Toyonaka 560-8531, Japan}

\date{\today}% It is always \today, today,

\begin{abstract}
We theoretically study the strain effects on the electronic properties of
 cobalt-based benzenehexathiol (CoBHT) coordination nanosheets using
 first-principles calculations. Two distinct crystal structures,
 high-density structure (HDS) and low-density structure (LDS), are
 explored. Our results reveal that HDS behaves as a metal, while LDS
 exhibits semiconducting. Spin-polarized electronic band structures highlight the presence of
energy band structures of Kagome lattice, and
 the inclusion of spin-orbit coupling (SOC) results in band gap openings
 at high-symmetric K points. Furthermore, we construct the tight-binding
 model to investigate the topological properties of CoBHT, 
 demonstrating anomalous Hall conductivity driven by the intrinsic
 Berry curvature. The impact of uniaxial 
 strain on the electronic and magnetic properties of CoBHT is also studied. Strain
 induces significant modifications in magnetic moments and density
 of states, particularly in the HDS. Anomalous Hall conductivity is
 enhanced under hole-doping conditions, suggesting that strain can be
 used to tailor the electronic properties of CoBHT for specific
 applications. Our findings underscore the potential of CoBHT nanosheets
 for use in next-generation electronic, optoelectronic, and catalytic
 devices with tunable properties through strain engineering. 
\end{abstract}

\maketitle

%\tableofcontents

%%%%%%%%%%%%%%%%%%%%%%%%%%%%%%%%%%%%%% section I %%%%%%%%%%%%%%%%%%%%%%%%%%%%%%%%%%%%%%
\section{\label{sec:intro}Introduction}
Two-dimensional (2D) materials, such as
graphene\cite{Novoselove2004.science,Novoselov.2005.nature,Novoselov.2005.PNAS,Geim.2007.naturemater}, 
boron-nitrides\cite{Watanabe.2004.naturemat,Li.2010.nanolett}, 
transition metal dichalcogenides (TMDCs)\cite{PhysRevLett.105.136805,10.1038/nnano.2010.279,10.1021/nl400516a}, and oxide nanosheets\cite{Osada2012.advmat}, have
garnered significant attention due to their unique physical and chemical
properties, i.e., spin and charge
transport\cite{10.1038/nature06037,10.1126/science.1250140,10.1038/ncomms16093,10.1021/nl4010783}, 
ferroelectric\cite{10.1021/nn102144s,10.1002/smll.201200752,10.1038/s41467-020-16291-9},
magnetism\cite{10.1038/nature22060,10.1038/nature22391},
and layer-by-layer oxidation\cite{10.1021/nl5049753,10.1021/acs.jpcc.8b05857}.
In further, the nontrivial topological properties of 2D materials
provide edge and corner states and optical shift current, which promise 
the electronic, spintronic, and quantum device applications\cite{RevModPhys.82.3045,PhysRevLett.95.146802,10.1088/2515-7639/ad2083,PhysRevB.107.115422,doi:10.1143/JPSJ.65.1920,PhysRevB.109.035431,10.1088/2053-1583/ab6ff7,PhysRevResearch.3.023121}.
These materials, often obtained via top-down exfoliation from bulk-layered
crystals, exhibit unique physical properties driven by their reduced
dimensionality. 
However, bottom-up approaches, where nanosheets are
synthesized through molecular, ionic, or atomic bonds, offer a
complementary method to tailor 2D material properties and create novel
structures. Coordination nanosheets
(CONASHs)\cite{Sakamoto2016,Sakamoto2017.Chemcomm,Kambe2013,10.1039/c9sc01144g}, a class of
2D materials composed of metal-organic frameworks, represent one such bottom-up
approach, enabling the design of nanosheets with versatile electronic,
magnetic, and optical characteristics.

The material properties of CONASHs can be fine-tuned 
by selecting different metal centers and ligands, covering a broad range of the
periodic table. 
The incorporation of transition metals into these
materials especially enhances their functionality.
Notable examples include the interfacial
synthesis of semiconducting nickel bis(dithiolene) (NiBHT)
nanosheets\cite{Kambe2013,Kambe2014.jacs}, photo-functional bis(dipyrrinato)zinc
nanosheets\cite{10.1038/ncomms7713}, and electrochromic iron or cobalt bis(terpyridine)
nanosheets\cite{10.1021/ja510788b}. Recent studies by Clough et al. further revealed the
potential of transition-metal dithiolene complex coordination polymers
in hydrogen evolution catalysis\cite{10.1021/ja5116937}. Theoretical work by Liu et al. has
predicted that single-layer NiBHT could function as a 2D topological
insulator\cite{Wang2013.nanolett}.

\begin{figure*}[hbt]
  \includegraphics[keepaspectratio, scale = 1.0]{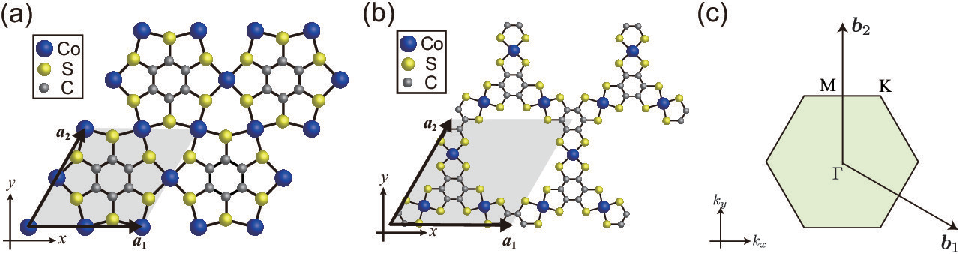}
  \caption
  {Crystal structures of CoBHT: (a) high-density structure (HDS) and
 (b) low-density structure (LDS). The gray rhombus is unit cell. CoBHT consists of Co (blue), S 
 (yellow), and C (gray) atoms. Here, $\bm{a_1}=(a,0)$ and
 $\bm{a_2}=a(1/2,\sqrt{3}/2)$ are primitive vectors, where $a$ is
 lattice constant. For HDS and LDS, $a=8.45$ and $14.52$\AA, respectively.
(c) The corresponding 1st Brillouin zone. 
 (BZ). Here, $\bm{b}_1=\frac{2\pi}{a}(1,-\frac{1}{\sqrt{3}})$ and 
$\bm{b}_2=\frac{2\pi}{a}(0,\frac{2}{\sqrt{3}})$.}
    \label{fig:structures} 
\end{figure*}

In this paper, we focus on the cobalt-based
coordination nanosheets (CoBHT)\cite{Pal2015}, constructed from cobalt, sulfur, and
carbon atoms, using density functional theory. the interfacial synthesis method.
Since CoBHT is a 2D thin film, applying uniaxial strain is
considered to significantly modify the electronic properties owing to
the mechanical flexibility,
similar to that observed in TMDCs, i.e.,
so-called strain
engineering\cite{Chaves2020,10.1021/nl4014748,10.1038/s41699-017-0013-7,10.1038/ncomms8381,10.1038/s41467-018-04099-7,Blundo2021}. 
%In further, it has been experimentally reported the 
%enhancement of hydrogen evolution from CoBHT under the external strain
%using the electrochemical spectroscopy technique\cite{kumatani,10.1021/acs.analchem.6b04355}.
%This result indicates a significant increase in electronic density of
%states (DOS) near the Fermi energy by the application of external strain.
Thus, in this paper, we study the effect of external strain on the 
electronic properties of CoBHT.
As previously reported, other transition metal-based BHT compounds have been experimentally synthesized in two crystal phases: high-density structure (HDS) and low-density structure (LDS), via a liquid-liquid interfacial reaction. CuBHT and FeBHT have been realized in the HDS phase\cite{Wang2021,10.1038/ncomms8408,wang2021two}, whereas NiBHT has been realized in the LDS phase\cite{Kambe2013}. Based on these experimental observations, we investigate the possibility that CoBHT could exist in both HDS and LDS phases owing to both posses a $D_{6h}$  symmetry with periodic arrangement of dithiolene groups. Clough et al. have reported Fourier-transform infrared (FTIR) spectroscopy of CoBHT in the LDS phase\cite{10.1021/ja5116937} and Li et al.  theoretically reported the HDS phase\cite{li2022first}. However, the precise crystal structure of CoBHT has not yet been experimentally confirmed.
Here we explore the electronic and topological properties of these two
different crystal structures using first-principles calculations
to identify the suitable crystal structure. 
Additionally, we
investigate how external strain influences the electronic and magnetic
behavior of CoBHT, offering insights into the tunability of these
properties for practical applications.

In Sec.II, we present the details of the crystal structure of CoBHT, and
study the electronic states for both HDS and LDS using first-principles
calculations. It will be shown that HDS becomes a spin-polarized
metallic state. However, LDS becomes a spin-polarized semiconductor
which is similar to NiBHT\cite{Wang2013.nanolett}. Energetically LDS is
slightly stable than HDS. Since Co atoms form Kagome lattice structure, the
energy band structures have Dirac cones and flat bands. It is also
pointed out that the spin-orbit interactions cause a small opening of
the energy band gap at Dirac cones. In Sec. III, we deduce the effective
tight-binding model by constructing the maximally localized Wannier
functions\cite{RevModPhys.84.1419,PhysRevB.56.12847} for CoBHT. Since
HDS has local magnetic moments originating from Co atoms, it shows an
anomalous Hall effect (AHE) owing to the finite Berry curvature. In
Sec. IV, we study the external strain effect on the electronic states of
CoBHT. 
The DOS for HDS near the Fermi energy is shown to be sensitive to the
external strain; however, there is almost no change for LDS due to its semiconducting nature.  
In Sec. V, we provide the summary of
the paper. In the Electronic Supplementary Information (ESI), we provide strain effects
of CoBHT on electronic band structures beyond $5$\% and simple
tight-binding analysis on charge density profile of
CoBHT.\dag

\section{Electronic States of C\lowercase{o}BHT}
CoBHT is a newly synthesized coordination
nanosheet composed of cobalt, sulfur, and carbon atoms.
Two crystal structures have been proposed for CoBHT, as shown in
Figs.~\ref{fig:structures}(a) and (b): HDS and LDS, respectively.
Both structures exhibit a periodic arrangement of
dithiolene groups and possess $D_{6h}$ symmetry. 
It should be noted that the Co atoms form a Kagome lattice structure in both cases.
The gray region represents the unit cell, and the
primitive lattice vectors for both structures are $\bm{a}_1 = (1, 0)a$
and $\bm{a}_2 = (1/2, \sqrt3/2)a$, with lattice constants of $a =
8.45$\AA\ for HDS and $a = 14.52$\AA\ for LDS.
The corresponding primitive vectors in reciprocal space are given as 
$\bm{b}_1=\frac{2\pi}{a}(1,-\frac{1}{\sqrt{3}})$
and 
$\bm{b}_2=\frac{2\pi}{a}(0,\frac{2}{\sqrt{3}})$. 
Therefore, the corresponding first Brillouin Zone
(BZ) is shown in Fig.~\ref{fig:structures}(c).

In this paper, the electronic structure calculations were performed with
the density functional theory (DFT)-based 
Quantum Espresso package using
the projector augmented wave pseudopotential method\cite{QE2017}. The exchange
correlation function was considered using the generalized gradient
approximation (GGA) by the Perdew-Burke-Ernzerhof (PBE)
method\cite{PBE}. We performed structural optimization of CoBHT using a variable cell relaxation procedure.
For structural optimization, the kinetic energy cut off 
was set to 85 Ry and the Brillouin zone 
was sampled a 12 $\times$ 12 $\times$ 1 grid 
based on a $\Gamma$-centered Monkhorst-Pack mesh\cite{PhysRevB.13.5188}.
The convergence threshold for the forces on each atom 
was set below $10^{-5}$ Ry/\AA, 
and the convergence criterion for the energy 
was set to $10^{-13}$ Ry.

\begin{figure*}[hbt]
\includegraphics[keepaspectratio, scale = 1.0]{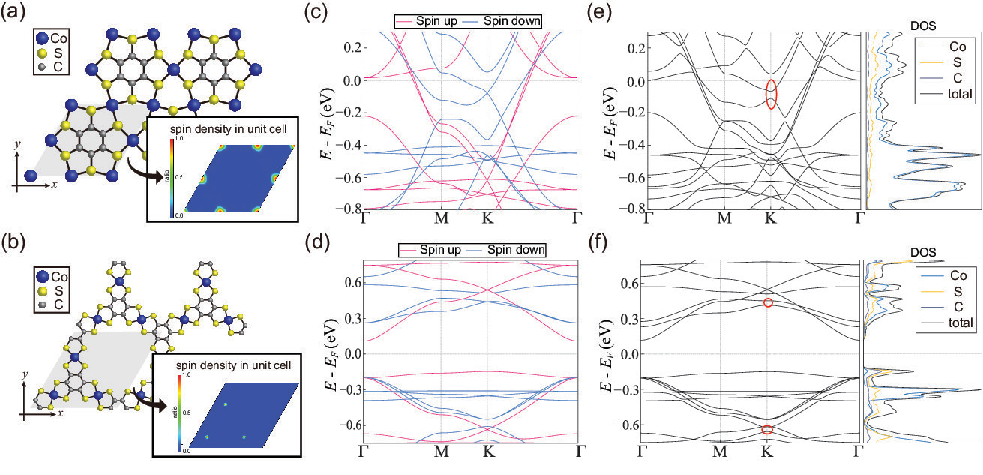}
\caption{Spin density plots of CoBHT in the unit cell for (a) HDS
and (b) LDS, respectively. CoBHT has finite magnetic moments
 originating from Co atoms. The magnetic moment of HDS is 1.85 $\mu_B$, while that of LDS is 4.07 $\mu_B$. 
Spin-polarized electronic band structures for (c) HDS and (d)
LDS, respectively. Red and blue lines represent up and 
down spin states, respectively. 
Figures (e) and (f) depict the electronic band
 structures and density of states (DOS) considering SOC for HDS and LDS, respectively. The band gap openings at the K point are marked by red ellipses. In the
 DOS plots, the blue lines represent the cobalt atoms, yellow lines
 represent sulfur atoms, gray lines represent carbon atoms, and black
 lines represent the total DOS. LDS exhibits
 semiconductor behavior with a band gap of $0.265$ eV. }
\label{fig:structure-band-dos} 
\end{figure*}

Figures~\ref{fig:structure-band-dos}(a) and (b) 
show the spin density plots within the unit cell for HDS and LDS,
respectively. Both HDS and LDS exhibit finite magnetic moments, which
originate from the cobalt atoms. The ground state of both systems exhibits ferromagnetic ordering. To verify this, we have compared the total energies of the systems in a ferromagnetic configuration and an antiferromagnetic configuration (in which one of the Co atoms is aligned oppositely to the other two). We have found that the ferromagnetic configuration has a lower total energy.
The magnitudes of these magnetic moments are 1.85 $\mu_B$ for HDS and
4.07 $\mu_B$ for LDS. The difference in the magnitude of magnetic moments between the HDS and LDS phases can be attributed to the role of the ligand, i.e., the different structural connectivity of Co atoms in the two phases (see Figs.1(a) and (b)). To confirm this, we calculate the Bader charges on each Co atom in both systems\cite{yu2011accurate}. We find that the average Bader charge on a Co atom is 8.29 (in arbitrary unit) in the HDS system and 8.25 (in arbitrary unit) in the LDS system. As a result, the magnitude of the magnetic moment differs between the two systems.
It is also worth comparing the binding energy, $\Delta E$, between HDS
and LDS. The binding energy is calculated using the following
equation: 
\begin{equation}
\Delta E = \frac{E_{total} -\sum_{i} N_i E_i}{\sum_{i}N_i}.
\end{equation}
Here, the summation index $i$ represents the Co, C and S atoms.
$E_{total}$ is the total energy of CoBHT for either HDS or LDS.
$N_i$ and $E_i$ denote the total number and total energy of $i$ atom,
respectively. 
The calculated values of $\Delta E$ are -0.5165 Ry/atom for LDS and 
-0.5065 Ry/atom for HDS, indicating that both HDS and LDS are
energetically stable. Moreover, LDS is slightly more stable than HDS
by 0.0010 Ry/atom. Further, we evaluated the thermodynamic stability using molecular dynamics simulations at 300 K, with the results presented in the ESI\dag\cite{andersen1980molecular}. We observed that both phases of CoBHT remained intact, preserving their hexagonal configurations and planarity compared to the 0 K structures. These findings confirm the thermal stability of both phases.

Figures~\ref{fig:structure-band-dos}(c) and (d) show 
the spin-polarized electronic energy band structure for HDS and LDS, respectively.
The red and blue lines indicate the spin-up and spin-down states, respectively.
The electronic energy band structure is calculated 
along highly symmetric directions in first BZ.
HDS is metallic. However, LDS becomes a semiconductor 
with a band gap of 0.265 eV.
Both structures exhibit linear dispersion at the $K$ point 
and feature a band structure resembling the Kagome lattice (Kagome-like band).
Additionally, we have calculated the electronic band structures by including an on-site Coulomb interaction (DFT+U) on each Co atom for both systems and present the results in the SM.\dag We find no changes in their magnetic ground state and electronic band dispersion. This result is consistent with previous theoretical study\cite{kang2022electronic}.
In SM, considering the simple tight-binding model of the Kagome lattice,
we compare the charge density plots obtained from DFT
with that obtained from the simple tight-binding model.\dag

Since CoBHT contains cobalt atoms, which induce relatively large intrinsic
spin-orbit interactions into the system, here we have taken 
spin-orbit coupling (SOC) into account.
Figures~\ref{fig:structure-band-dos}(e) and (f) depict 
the electronic energy band structures with SOC,
together with the partial density of states (PDOS) for HDS and LDS, 
respectively.
In both structures, finite SOC opens the energy band gap at K point (as marked by the red ellipses). 
Furthermore, from the PDOS near the Fermi energy, 
it is evident that in HDS the contribution of cobalt electrons is predominant, 
while in the LDS, there is a contribution not only from cobalt electrons but also 
from sulfur and carbon electrons.
%%%In subsection III, we discuss the topological properties 
%%%of HDS using the effective tight-binding model Hamiltonian.

\begin{figure*}[hbt]
\centering
\includegraphics[keepaspectratio, scale = 1.0]{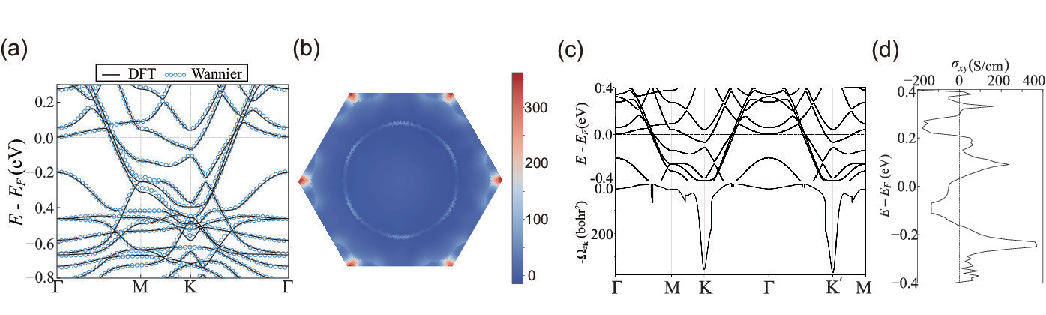}
\caption{(a) Comparison of the band structures of CoBHT for HDS with
 SOC as calculated by DFT (the black line) and the WTB Hamiltonian (the
 blue circles). The WTB Hamiltonian consists of all $d$-orbitals of the
 cobalt atoms. (b) Contour plot of Berry curvature ($-\Omega_{z
 \bm{k}}$) in the first BZ. (c) Energy band structure and
 corresponding Berry curvature ($-\Omega_{z\bm{k}}$) along the path
through the high-symmetric points in the first BZ.
(d) Fermi energy dependence of anomalous Hall conductivity of CoBHT for HDS.
}
\label{fig:WTBM-DFT-Berry}
\end{figure*}

%%%%%%%%%%%%%%%%%%%%%%%%%%%%%%%%%%%%%% section III-A %%%%%%%%%%%%%%%%%%%%%%%%%%%%%%%%%%%%%%
\section{Tight-Binding Model and Berry curvature}
In order to analyze the topological properties of CoBHT, we shall
construct the effective tight-binding model of HDS using Wannier90\cite{Wannier90} to
reproduce the energy band structure obtained by DFT.  
In HDS, the electronic states near the Fermi energy is predominantly occupied 
by cobalt electrons. 
Furthermore, since $d$-orbitals of cobalt electrons contribute 
in this energy range, we consider all $d$ orbitals ($d_{xy}$, $d_{xz}$,
$d_{yz}$, $d_{z^2}$, $d_{x^2-y^2}$) for each of the three cobalt atoms in the unit cell, 
i.e., totatally 15 orbitals.
The size of the effective tight-binding Hamiltonian is 30 $\times$ 30,
because the SOC derived from cobalt atoms 
is also considered.

Figure~\ref{fig:WTBM-DFT-Berry}(a) shows the energy band 
obtained by DFT and the effective model 
for HDS obtained by Wannier90. The black lines represent the energy band
dispersion from DFT, 
while the blue circles represent the energy band dispersion
from the effective tight-binding model. Thus, the energy band structure of HDS of
CoBHT can be well-described by the $d-$orbitals of cobalt atoms.
Since the cobalt atoms form a 2D katome lattice, 
further analysis using simple tight-binding model is presented in SM.\dag

Since HDS has a spin-polarized metallic state owing to the local magnetic
moment of Co atoms, it is expected that HDS exhibits
AHE\cite{Nagaosa2010.RMP,Yao2004.PRL,RevModPhys.82.1959,PhysRevB.74.195118,SMIT1955877,SMIT195839,PhysRevB.2.4559}, 
i.e., finite Hall conductivity without an external magnetic field. 
It is known that there are two main mechanisms for AHE, i.e., extrinsic and
intrinsic mechanisms. 
The extrinsic one is attributed to the skew
scattering\cite{SMIT1955877,SMIT195839} or
side-jump\cite{PhysRevB.2.4559} from disorder. 
The intrinsic one can be attributed to the topological
properties of bulk wave functions, which occur even in the perfect
crystal.
Here, we shall focus on the intrinsic AHE.
The anomalous Hall conductivity can be obtained by the
$\bm{k}$-integration of the Berry curvature in the 1st BZ as
\begin{eqnarray}
  \sigma_{xy} 
  = 
  -\frac{e^2}{\hbar} \int_{\text{BZ}} \frac{d^3k}
  {(2 \pi)^3} \Omega_{z \bm{k}}.
  \label{eq:AHC}
\end{eqnarray}
Here $\Omega_{z \bm{k}}$ is the summation of Berry curvature upto the
Fermi energy, which is given as
\begin{eqnarray}
  \Omega_{z \bm{k}} 
  = 
  \sum_n f_n \Omega_{zn \bm{k}},
\end{eqnarray}
where 
$f_n$ is the Fermi-Dirac distribution function and $n$ is the band index.
The Berry curvature for $n$-th band can be numerically evaluated through
the Kubo formula, i.e., 
\begin{eqnarray}
  \Omega_{zn \bm{k}} 
  = 
  -\sum_{n^\prime \ne n} \frac{2\text{Im} 
  \braket{\Psi_{n \bm{k}}|v_x|\Psi_{n^{\prime} 
  \bm{k}}}\braket{\Psi_{n \bm{k}}|v_y|\Psi_{n^{\prime} 
  \bm{k}}}}{(\omega_{n^{\prime}} - \omega_n)^2},
\label{eq.bc}
\end{eqnarray}
where $v_{x(y)}$ is the $x(y)$ component of the velocity operators, $\omega_n = E_n/\hbar$.

Since CoBHT is ferromagnetic, i.e., a time-reversal broken system, the Berry curvature 
has the property of $\Omega_{z,\bm{k}} = \Omega_{z,-\bm{k}}$.
Figure~3(b) shows the contour plot of Berry curvature
$\Omega_{z,\bm{k}}$ in the 1st BZ.
Figure~3(c) shows the energy band structure of HDS near the Fermi energy
and the corresponding Berry curvature
$\Omega_{z,\bm{k}}$ along the high symmetric $\bm{k}$ points of 1st BZ.
The Berry curvature of HDS clearly 
exhibits a six-fold rotational symmetry with respect to the $\Gamma$ point.
It is clearly seen that the most pronounced peaks appear at $K$ and
$K^\prime$ points, where Dirac points exist owing to the nature of 2D
Kagome lattice.
Thus, as shown in Fig.~3(d), the finite anomalous Hall conductivity
$\sigma_{xy}$ is obtained for CoBHT, which is a profound value.

%%%%%%%%%%%%%%%%%%%%%%%%%%%%%%%%%%%%%% section IV %%%%%%%%%%%%%%%%%%%%%%%%%%%%%%%%%%%%%%
\section{Strain Effect on C\lowercase{o}BHT}
In atomically-thin 2D materials, the electronic states can be
significantly modified by the application of external
strain~\cite{10.1038/s41377-020-00421-5,10.1021/nn800459e,Choi2010.prb}. Here
we study the strain effect on electronic states of CoBHT. 
Since CoBHT has $D_{6h}$ hexagonal symmetry, 
the application of uniaxial strain breaks the hexagonal crystal symmetry,
resulting in the significant modification of electronic states.
Furthermore, recent studies show that the strain can induce the
topological transition in 2D Kagome lattice~\cite{Mojarro_2024}. 
\begin{figure*}[hbt]
  \centering
  \includegraphics[keepaspectratio, scale = 1.0]{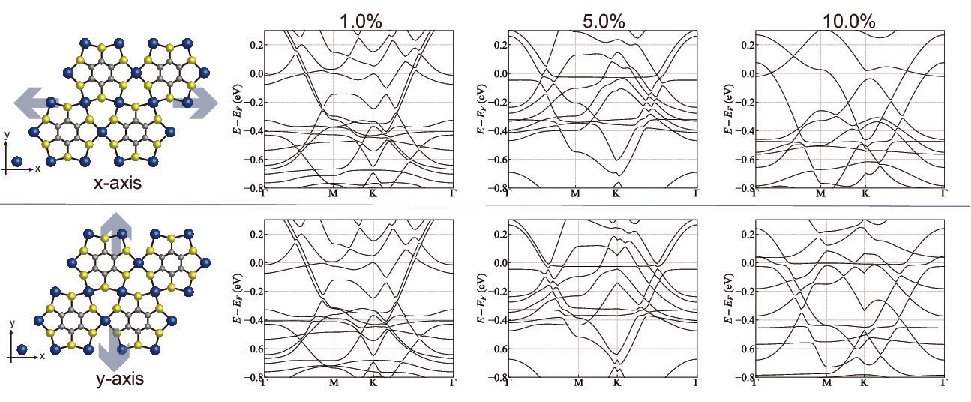}
  \caption{Strain effect of CoBHT for HDS on the electronic energy band structures.
(Upper panels) Elongation strain along $x$-axis with the strain $1.0$, $5.0$,
 and $10.0$\%. (Lower panels) Elongation strain along $y$-axis.}
\label{fig:anisotropy-band}
\end{figure*}

Figure~\ref{fig:anisotropy-band} shows the energy band structures of HDS
under the application of external strains for $1.0$\%, $5.0$\%, and
$10.0$\%, respectively. Here, the elongation strain is applied. The
effect of compression strain and weaker strain less than $1.0$\% is
presented in SM.\dag
The upper and lower panels of Fig.~\ref{fig:anisotropy-band} correspond
to the external strain along $x$ and $y$ axes, respectively. 
Since applying the strain to HDS breaks
the hexagonal symmetry of the system, 
a gap opens up in the linear dispersion 
at the K point.
\begin{figure*}[hbt]
  %figure 5
  \centering
  \includegraphics[keepaspectratio, scale = 1.0]{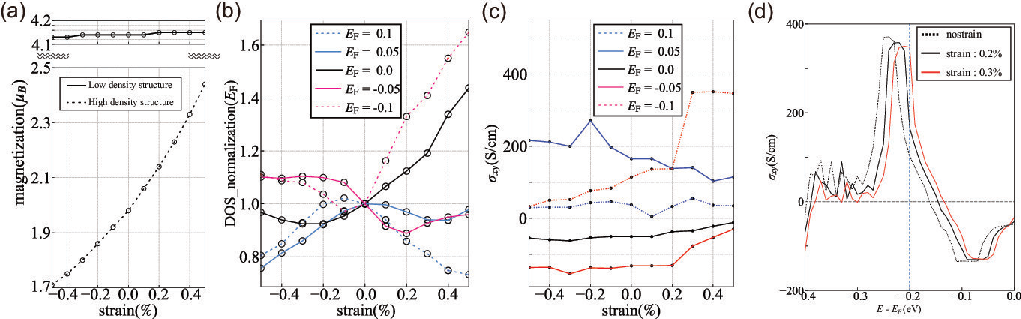}
  \caption{(a) Strain dependence of magnetic moment for (lower) HDS and
 (upper) LDS, respectively. Here $E_F$ is fixed at $0$ eV, i.e.,
 non-doping case. (b) The strain effect of DOS at $E_F$ for HDS is for
 several different electron or hole doping cases. 
(c) Strain effect of anomalous
 Hall conductivity of HDS. (d) Fermi energy dependence of anomalous Hall
 conductivity for HDS with several different strains.
  }
\label{fig:magdos}
\end{figure*}

It is also observed that strain induces the anisotropic effect 
in the electronic states of CoBHT, which becomes sizable
in the range of $5.0$\% to $10.0$\%.
%In the experiment, the maximum strain is estimated at about $0.38$\%\cite{kumatani}.
However, the following discussion shall focus on the uniaxial strain up to $0.5$\%, within a experimental perspective.
In case of biaxial strain, the further large strain can be achieved
using indenting devices~\cite{10.1126/science.1157996,10.1021/acs.nanolett.8b03833,10.1038/ncomms15815}.

Figure~\ref{fig:magdos}(a) illustrates the strain dependence of the magnetic
moment for HDS and LDS of CoBHT at the charge-neutral point, i.e., $E_F=0$.
For HDS, the magnetic moment monotonically increases with increase in strain.
However, for LDS, the magnetic moment remains nearly
unchanged. Thus, the HDS has a stronger strain dependence of magnetic
moments than the LDS. Since the magnetic 
moments originate from the $d$-orbitals of Co atoms, these results
indicate that the strain affects the spin-spin interactions
in HDS more than LDS.
This might be attributed to the fact that the atomic distances between
Co atoms differ significantly between HDS and LDS. 
In other words, HDS (LDS) has the stronger (weaker) magnetic interactions between Co atoms.

Figure~\ref{fig:magdos}(b) 
shows the strain dependence of DOS for HDS
at several different Fermi energies.
The values of DOS are normalized by the DOS at $E_F =0$ with no strain.
At the charge neutral point, applying a $0.5$\% strain 
results in an approximately $40$\% increase in the DOS 
compared to the case without strain.
Taking the doped case into consideration, 
it was found that applying a $0.5$\% strain, 
especially in the hole-doped case ($E_F = -0.1$ eV), 
results in an approximately $60$\% increase.
One anticipated application of CONASHs is their utilization 
as electrode catalyst nanosheets, and the results suggest 
the possibility of activating catalytic functions.
The LDS has a band gap of $0.265$ eV, 
so there is no DOS in the range where $E_F$ is from $-0.1$ to $0.1$.

Figure~\ref{fig:magdos}(c) depicts the anomalous Hall conductivity 
of HDS obtained by applying strain 
and calculating it at several different Fermi energy levels, 
similar to the procedure employed for DOS calculations.
The calculation of anomalous Hall conductivity
was performed using Eq.~(\ref{eq:AHC}) presented in Sect.~III.
The change in conductivity due to strain is generally small 
on average, but for the hole-doped case with $E_F = -0.2$, 
a significant increase was observed when applying strain from $0.2$\% to $0.3$\%.

Figure~\ref{fig:magdos}(d) shows graphs for cases where the strain of $0.2$\% and $0.3$\% was applied, 
with the horizontal axis representing Fermi energy 
and the vertical axis indicating conductivity.
The blue dashed line shows $E_F = -0.2$.
Upon comparing each graph, 
it can be observed that when applying strain from $0.2$\% to $0.3$\%,
the peaks near the blue dashed line precisely overlap.
Therefore, when strain is applied at $0.3$\% or higher, 
the conductivity takes on significant values.

\section*{Conclusions}
In this paper, we investigated the electronic structure and strain
effects of CoBHT using first-principles calculations. Our results demonstrate that CoBHT
exhibits diverse electronic and magnetic properties, with the
HDS showing metallic behavior and the
LDS functioning as a semiconductor with a band
gap of 0.265 eV. The inclusion of SOC further
reveals a band gap opening at the K points, contributing to the
topological properties of the system. We also analyzed the Kagome-like
band structure and anomalous Hall conductivity using the Wannier
tight-binding model, confirming the non-trivial topological nature of
the HDS. 

Furthermore, we explored the strain effects on the electronic properties
of CoBHT, showing that uniaxial strain can induce significant changes in
magnetic moments, DOS, and anomalous Hall
conductivity. These findings suggest that strain engineering could be a
viable approach to enhance the electronic functionality of CoBHT,
particularly in applications requiring tunable electronic, magnetic, and
catalytic properties. This work underscores the potential of CoBHT
nanosheets for next-generation electronic and optoelectronic devices, as
well as advanced catalytic applications.

\section*{Acknowledgements}
The authors are grateful to A. Kumatani for helpful discussions and
showing us his experimental data in advance of publication. 
This work was supported by JSPS KAKENHI (Grants No. JP25K01609,
No. JP22H05473, and No. JP21H01019), JST CREST (Grant
No. JPMJCR19T1). K. W. acknowledges the financial support for Basic
Science Research Projects (Grant No. 2401203) from the Sumitomo
Foundation.  

%%%END OF MAIN TEXT%%%


%If notes are included in your references you can change the title
%from 'References' to 'Notes and references' using the following
%command: 

%\nocite{*}

\bibliography{reference}% Produces the bibliography via BibTeX.

\end{document}